\documentclass[review]{elsarticle}

\usepackage{hyperref}

\usepackage{subcaption}
\usepackage{graphicx}
\usepackage{grffile}

\journal{Journal of \LaTeX\ Templates}

\usepackage{siunitx}

\def\DH#1#2{^{#2}_{\Lambda\Lambda}{\rm #1}}

\def\micron{\si{\micro\metre}}

\begin{document}

\begin{frontmatter}

\title{CNN-based event classification for alpha-decay events in nuclear emulsion}


\author[tohoku,riken]{J. Yoshida\corref{mycorrespondingauthor}}
\cortext[mycorrespondingauthor]{Corresponding author}
\ead{jyoshida@lambda.phys.tohoku.ac.jp}

\author[riken]{H. Ekawa}
\author[riken,gifu]{A. Kasagi}
\author[riken]{M. Nakagawa}
\author[gifu]{K. Nakazawa}
\author[riken]{N. Saito}
\author[riken,gsi,lanzhou]{T.R. Saito}
\author[rikkyo]{M. Taki}
\author[gifu]{M. Yoshimoto}


\address[tohoku]{Physics Department, Tohoku University, Aramaki, Aoba-ku, Sendai 980-8578, Japan}
\address[riken]{High Energy Nuclear Physics Laboratory, Cluster for Pioneering Research, RIKEN, 2-1 Hirosawa, Wako, Saitama 351-0198, Japan}
\address[gifu]{Physics Department, Gifu University, 1-1 Yanagido, Gifu 501-1193, Japan}
\address[gsi]{GSI Helmholtz Centre for Heavy Ion Research, Planckstrasse 1, D-64291 Darmstadt, Germany}
\address[lanzhou]{School of Nuclear Science and Technology, Lanzhou University, 222 South Tianshui Road, Lanzhou, Gansu Province, 730000, China}
\address[rikkyo]{Graduate School of Artificial Intelligence and Science, Rikkyo University, 3-34-1 Nishi Ikebukuro, Toshima-ku, Tokyo 171-8501, Japan}

\begin{abstract}
We developed an efficient classifier that sorts alpha-decay events from various vertex-like objects in nuclear emulsion using a convolutional neural network (CNN).
Alpha-decay events in the emulsion are standard calibration sources for the relation between the track length and kinetic energy in each emulsion sheet.
We trained the CNN using 15,885 images of vertex-like objects including 906 alpha-decay events and tested it using a dataset of 46,948 images including 255 alpha-decay events.
By tuning the hyperparameters of the CNN, the trained models achieved an Average Precision Score of 0.740 $\pm$ 0.009 for the test dataset.
For the model obtained, a discrimination threshold of the classification can be arbitrarily adjusted according to the balance between the precision and recall.
The precision and recall of the classification using previous method without a CNN were 0.081 $\pm$ 0.006 and 0.788 $\pm$ 0.056, respectively, for the same dataset.
By contrast, the developed classifier obtained a precision of 0.547 $\pm$ 0.025 when a similar recall value of 0.788 was set.
The developed CNN method reduced the human load for further visual inspection after the classification by approximately 1/7 compared to the estimated load of the former method without a CNN.


\end{abstract}


\end{frontmatter}


\section{Introduction}

Nuclear emulsion is one of the detectors used for visualising the tracks of charged particles with the highest spatial resolution at the micrometre-scale or better \cite{Barkas}.
Such an excellent spatial resolution has provided numerous opportunities in fundamental studies \cite{LATTES1947, 10.1143/PTP.46.1644, 10.1143/PTP.85.1287, KODAMA2001218, 10.1093/ptep/ptu132} and applications \cite{TANAKA2007104, Morishima2017} during the past 80 years.

One of the recent areas to employ the emulsion is the experimental investigation into double hypernuclei, which are baryonic bound states with two strange-quarks \cite{PhysRevC.88.014003, E07}.
Studies on double hypernuclei have extended our understanding of the nuclear force to the general baryon-baryon interaction under the flavoured-SU(3) symmetry.
Double hypernuclei are produced through the capture of a $\Xi ^-$ hyperon with two strange-quarks in a nucleus.
In the double strangeness system produced, a conversion process of $\Xi^-$p$\rightarrow \Lambda \Lambda$ takes place, followed by the decay. Because of the small Q-value of the conversion process, i.e., approximately 28 MeV, 
particles and fragments from the production and decay of the double hypernuclei have small kinetic energy, and the track length of these particles is extremely short, i.e., typically on the order of 10 micrometres in a solid material. 
When a $\Xi ^-$ hyperon is stopped in the emulsion, visual information of the tracks can be recorded simultaneously with the production and decay of the double hypernuclei. Through visual analyses of the length and boldness of the recorded tracks, particles  and fragments are identified and their kinetic energy can be deduced.
Therefore, the produced double hypernucleus can be identified and its mass value can be obtained even with only one event observed in the emulsion. 



Experimental studies on double hypernuclei using the emulsion have made significant progress during the last decade.
The most recent experiment for studying double hypernuclei was carried out in J-PARC as the E07 experiment \cite{E07}.
The basic design of the experiment conducted was based on the hybrid emulsion method. In this method, only a small area of the emulsion sheet is scanned, and the area is defined by tracking information of the $\Xi ^-$ particle measured by the other precise detectors in front of the emulsion. This drastically reduces the load and time for the analysis. 
Owing to the high beam intensity, large solid angle of the detectors, and modern techniques used in an emulsion analysis \cite{MYINTKYAWSOE201766}, 10-times more double hypernuclear events have been expected to be detected than those observed in the former experiments \cite{PhysRevC.88.014003}.
The experimental data are being analysed and several nuclei, including $\DH{Be}{}$, have been identified thus far \cite{ekawa2019}.
However, the performance of the hybrid method has yet to be perfected, and only one-third of the expected candidates have been observed. 

An exhaustive search method as an alternative to the hybrid method is also being developed \cite{YOSHIDA201786}, and the first $\Xi$ hypernucleus candidate was observed \cite{10.1093/ptep/ptv008, doi:10.1146/annurev-nucl-101917-021108}.
With this method called "overall scanning" herein, the entire volume of the irradiated emulsion is scanned, and therefore the method is capable of detecting events of a non-triggered double hypernuclear production. With the overall scanning of emulsion sheets irradiated during the J-PARC E07 experiment, the detection of approximately $10^3$ events related to the production and decay of double hypernuclei is expected. 
Although the load for the analysis and the amount of scanned data will be drastically increased with the existing  overall scanning technique, the technique is expected to take on the main role in studies on double hypernuclei and could replace the presently used hybrid method.

\begin{figure}[t]
\begin{center}
 \begin{minipage}[b]{0.2\linewidth}
  \centering
  \includegraphics[keepaspectratio, scale=0.22]
  {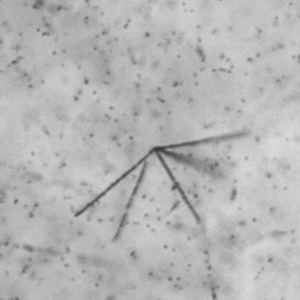}
  \subcaption{}
 \end{minipage}
 \begin{minipage}[b]{0.2\linewidth}
  \centering
  \includegraphics[keepaspectratio, scale=0.22]
  {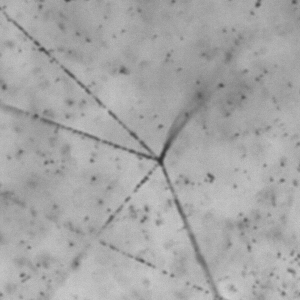}
  \subcaption{}
 \end{minipage}
 \begin{minipage}[b]{0.2\linewidth}
  \centering
  \includegraphics[keepaspectratio, scale=0.22]
  {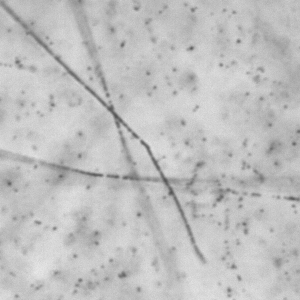}
  \subcaption{}
 \end{minipage}
 \begin{minipage}[b]{0.2\linewidth}
  \centering
  \includegraphics[keepaspectratio, scale=0.22]
  {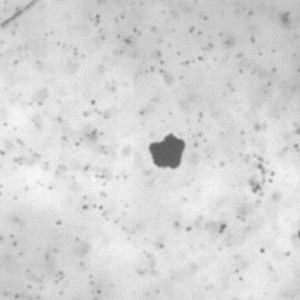}
  \subcaption{}
 \end{minipage}
 \end{center}
 \caption{Images of typical objects detected using Vertex Picker. (a) An alpha-decay event of a thorium series, (b) interaction of a hadron beam particle and a nucleus in an emulsion layer, (c) cross of unrelated tracks, (d) a black spot similar to dust in an emulsion layer. The sizes of these images are 224 pixels $\times$ 224 pixels and 87 $\times$ 87 $\micron^2$ at the object.}\label{fig:objects}
\end{figure}

For the overall scanning approach, we previously developed a scanning system called "Vertex Picker" \cite{YOSHIDA201786}.
This system takes exhaustive micro-graphs of thick emulsion sheets and detects vertex-like objects with three or more tracks originated from a single point.   
However, this method detects not only the vertex candidates of hypernuclei but also a large number of other vertex-like objects (e.g., an alpha-decay event and a beam-nucleus interaction) and non-vertex objects (e.g., a cross of unrelated tracks and a black spot similar to dust), as shown in Figure \ref{fig:objects}. 
The developed system improves the speed of the vertex-search by a factor of approximately 20; however, the speed must be further improved for the overall scanning technique.
Moreover, the ratio of the detected vertex events of interest to the other detected events is far from  satisfactory.

To achieve further improvements, we developed a new technique for detecting vertex-like objects in the emulsion by employing a convolutional neural network (CNN).
Image classification using a CNN has made remarkable progress in recent years and has reached a level comparable to that of human visual classification \cite{LeCun2015, DBLP:journals/corr/HeZR015}.

Our ultimate goal is to develop a fast method for detecting candidates of vertices related to double hypernuclei with a large detection efficiency and an excellent signal-to-background ratio. However, there are insufficient amounts of training and validation data on double hypernuclei for the developmentof a method using a CNN. Therefore, as the first step, we attemped to develop a method using a CNN for detecting alpha-decay events in the emulsion. 
Alpha-decay events are traces of a spontaneous chain decay of long-lived radioisotopes such as uranium and thorium in the emulsion.
Recorded images of alpha-decay events of uranium and thorium series have been characterised as consisting four and five bold tracks of approximately 25-50 $\micron$ in length, respectively.
Those alpha-decay events and associated tracks are extremely important because we use them for calibrating the relation between the track length and kinetic energy in each emulsion sheet.
The present study employs a sufficient number of alpha-decay events selected by Vertex Picker and by the human eye during the development using a CNN. 
Furthermore, this study attempts to transition the emulsion techniques from a mature approach to a state-of-art technology by introducing a CNN, and the present work is a foundation for the further developments towards studies of double hypernuclei with the emulsion.

\section{Event classification using CNN}

\subsection{Developed method}
For the development of an event classifier for alpha-decay events using a CNN, we prepared three datasets for training (TRAIN), validation (VALID), and testing (TEST). Prior to the development, we already had images selected from six of the 50 mm $\times$ 50 mm $\times$ 0.5 mm volumes of two emulsion sheets irradiated during the J-PARC E07 experiment, which are used for the TRAIN and VALID datasets.
A total of 1120 images of alpha-decay events and 18,793 images of other events are randomly distributed in the TRAIN and VALID datasets at a ratio of 4:1. 
Furthermore, we scanned another volume of an additional emulsion sheet, and 255 alpha-decay images and 46,693  images of other events were used as the TEST dataset. 
A summary of the dataset is shown in Table \ref{table:dataset}.
All images were sorted using Vertex Picker and the human eye.

\begin{table}[t]
\caption{Summary of dataset and numbers of images.}
\label{table:dataset}
\begin{center}
\begin{tabular}{l c c c}
\hline
Dataset name& Alpha-decays & Others  & Total\\
\hline
Training (TRAIN) &	906    &	14979 	&  15885\\
Validation (VALID) & 214    &	3814  &	 4028 \\
Test (TEST) &  255    &	46693 &  46948 \\
\hline
\end{tabular}
\end{center}
\end{table}

We employed a ResNet50 CNN, which has already shown excellent results in competitions on large-scale image classification \cite{HeZRS15}.
The inputs of the CNN are colour images, and the original output is a set of probabilities for 1000 categories of objects.
We modified ResNet50 to output a scalar value for each image by adding three stages of fully connected layers at the end part of the CNN instead of the 1000-channel layer.
The output shall be a positive value for an alpha-decay event and a negative value for another object.
The threshold of the discriminant function used to distinguish between  positive and negative samples is usually 0.0; however, 
we varied the threshold value to study the performance of the event classifier, as discussed in section 2.2.

The programs for this training are written in Python 3.6.9, PyTorch 1.5.1  \cite{paszke2017automatic}, and PyTorch Lightning 0.8.5 \cite{falcon2019pytorch}.
Because the ratio of alpha-decay images to other images is approximately 1:17 in the TRAIN and VALID datasets, this imbalance may affect the training process. To overcome this problem, we combined two techniques, data augmentation and over sampling. During the training, there are 96 images in each mini-batch process, and they are randomly selected from the TRAIN dataset.
To create the balance, the selection probability for the alpha-decay data is set to approximately 17-times larger than the others during training, which is a so-called over sampling method \cite{DBLP:journals/corr/abs-1710-05381, imbalanced-dataset-sampler}.
Furthermore, to provide a generalisation to the developed CNN, the training data must be diverse, and we therefore also introduced a random flip and RandAugment \cite{cubuk2019randaugment, pytorch-randaugment} as the data augmentation techniques.

The RandAugment executes several image transformations automatically and randomly by specifying two hyperparameters, which are defined as N and M.
Parameter N is the number used to specify how many image transformations are randomly selected from the defined transformations.
In this study, we employed eight types of image transformation: rotation, change in contrast, change in brightness, change in sharpness, transformation into a parallelogram tilted in the horizontal and vertical directions, and a shift in the horizontal and vertical directions.
The other parameter M is the magnitude of the image transformations within the range of 0-30, where a value of 30 corresponds to a magnitude of 100 \%.
According to the original study on RandAugment, the optimal value of (N, M) varies depending on the dataset and the structure of the CNN.
Therefore, in the present development, the performance of the trained model was studied by varying the parameters N and M, the details of which are discussed in the next sub-section.

%

Because the number of images used in our development is insufficient to train the CNN from a random state, we employed a CNN model already pre-trained with large-scale data, i.e., the ImageNet dataset, as the initial state \cite{5206848}. 
During the training process, the size of the mini-batch is set to 96, which is the maximum value that fits the GPU  applied.
To optimise the weight of the convolutions in the CNN, we used the Adam optimizer \cite{kingma2014adam}, which has been widely used in other developments of a CNN. As a loss function, we employed binary cross entropy, which is considered one of the best loss functions for binary classifications.



\subsection{Comparison metrics}

As discussed in the previous sub-section, we have to find an (N, M) parameter combination to achieve the optimal CNN performance to efficiently classify alpha-decay events from other events.
To evaluate the performance of the classifiers, we use the area under the Precision-Recall Curve, also known as the Average Precision Score" \cite{10.1145/1143844.1143874}.
A Precision-Recall Curve is widely used to visualise the performance of a binary classifier, particularly for an imbalanced dataset like ours.
The curve consists of pairs of two parameters, widely referred to as precision and recall at different threshold values to discriminate positive and negative samples based on the output values of a CNN.
The precision corresponds to the purity in the classified samples, whereas the recall represents the selection efficiency of the classifier.


The procedure for selecting the best (N, M) is as follows.
Initially, we conduct the training using a specified (N, M) until the minimum validation loss is observed, and we defined the best model having the minimum validation loss.  
After the training, we evaluated the Average Precision Score for the VALID dataset.
We applied this process four times with the same (N, M) and different random seeds to check the reproducibility.
We iterated this process with various pairings of (N, M), and the pair considered as the best occurs when the mean of the four Average Precision Scores is at maximum.
Finally, to evaluate the averaged performance, we applied the best four models individually to the TEST dataset with the chosen (N, M). 



%

\begin{figure}[t]
 \centering
 \includegraphics[keepaspectratio, scale=0.4]
 {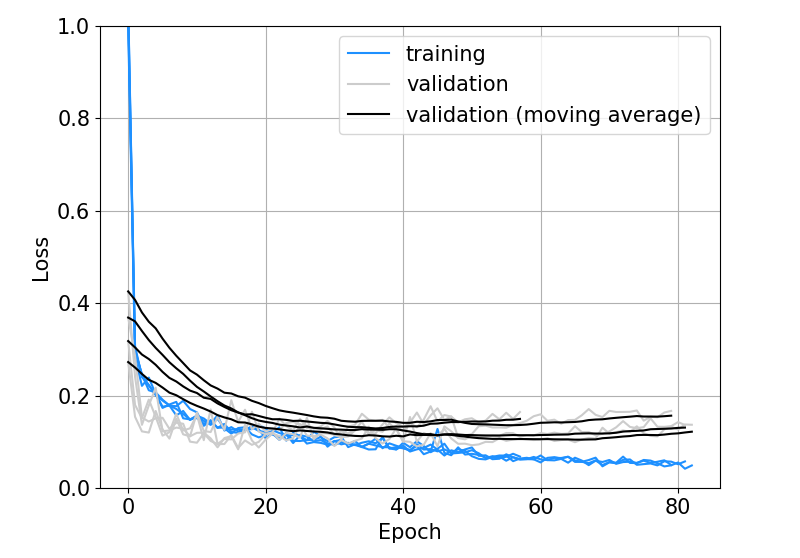}
 \caption{Evolution of the losses for the TRAIN and VALID dataset for four iterations of the training process at (N = 2, M = 24), as an example. The best epoch for each of the four trials was 36, 54, 55, and 65, respectively.}
 \label{fig:aps_curve}
\end{figure}

\section{Results of the trained CNN model}

We conducted a grid search of the hyperparameters of RandAugument, i.e., N and M, for N = \{2, 4, 6, 8\} and M = \{6, 12, 18, 24, 30\}. 
For each (N, M) pair, we searched the best model at the specific epoch providing the minimum loss for the VALID dataset. 
Figure \ref{fig:aps_curve} shows the evolution of losses for four iterations of the training process, for example, at (N = 2, M = 24). 
The blue and grey lines represent the loss values, respectively, of the TRAIN and VALID dataset for the four trials. 
The loss values of the TRAIN dataset decrease as the training progresses; however, the loss of validation begins to increase gradually at a certain point. 
As described in the previous section, we define the best model as having the minimum validation loss. To find the epoch number providing the minimum validation loss, we applied a smoothing for the validation loss values with a method called the exponentially weighted moving average, which is formulated through the following recurrence formula:
\begin{equation}
\begin{array}{l}
S_0 = V_0, \\
S_i = wS_{i-1}   +  (1.0 - w) V_i,
\end{array}
\end{equation}
where $S_i$ and $V_i$ are the $i$-th smoothed value and the original value, respectively, and $w$ is a weight taking a value of between zero and 1 and specifies the degree of smoothing.
In the present study, we set the weight to 0.9 and stopped training when the smoothed loss exceeded 115\% of the minimum.

\begin{figure}[t]
 \centering
 \includegraphics[keepaspectratio, scale=0.4]
 {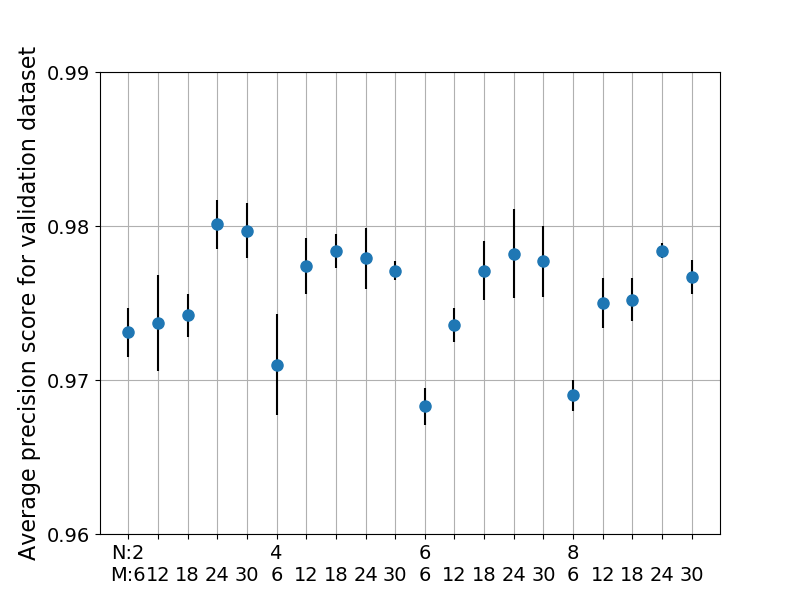}
 \caption{The Average Precision Score for the VALID dataset for various pairings (N, M) for RandAugment. The best value is 0.980 $\pm$ 0.002 at (N = 2, M = 24).}
 \label{fig:NMsearch}
\end{figure}

Through a grid search among N = \{2, 4, 6, 8\} and M = \{6, 12, 18, 24, 30\}, we obtained the mean and standard deviation of the four Average Precision Scores for the VALID dataset, as shown in Figure \ref{fig:NMsearch}.
The best combination is achieved with (N = 2, M = 24), and the best score obtained is 0.980 $\pm$ 0.002.

\begin{figure}[t]
 \centering
 \includegraphics[keepaspectratio, scale=0.4]
 {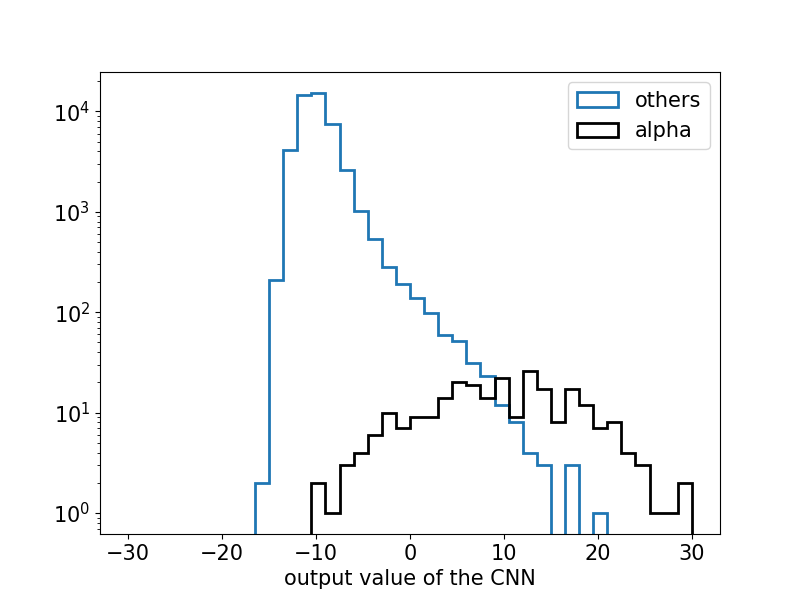}
 \caption{Distribution of the output values of one of the best CNN models at (N = 2, M = 24) for alpha-decay and other events in the TEST dataset.}
 \label{fig:logits}
\end{figure}
\begin{figure}[t]
 \centering
 \includegraphics[keepaspectratio, scale=0.4]
 {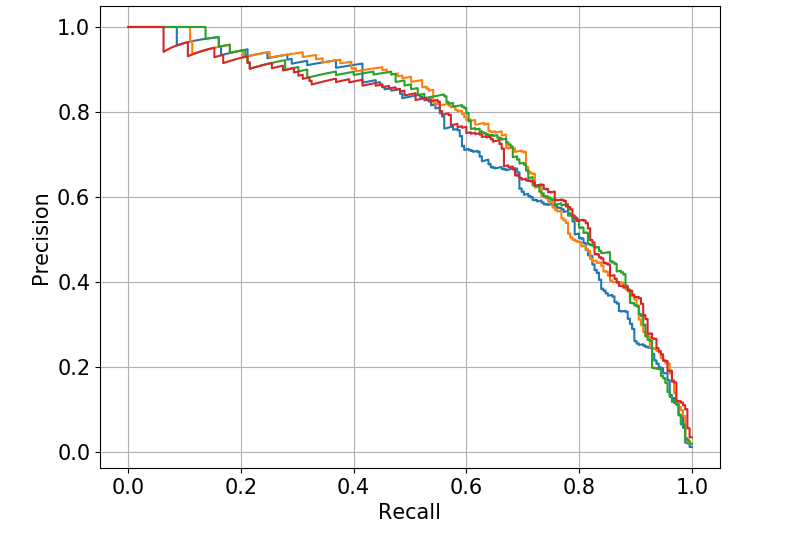}
 \caption{Precision-Recall curves at (N = 2, M = 24). This curve consists of pairs of precision and recall at different discriminant threshold values for the output of the CNN.}
 \label{fig:PR}
\end{figure}

We applied the classification of the alpha-decays in the TEST dataset with the four selected models at the best pairing (N = 2, M = 24).
Figure \ref{fig:logits} shows the distribution of output values of one of the four CNN models for the alpha-decay  and other events.
The values for the non-alpha-decay events are distributed mainly in the negative region, whereas, the distribution of the output values for alpha-decays is shifted to positive values. 
Figure \ref{fig:PR} shows the four Precision-Recall Curves representing the correlation of precision and recall by varying  the discriminant threshold value.
The Average Precision Score, i.e., the area under the curve, is 0.743 $\pm$ 0.007.


\section{Comparison with former method without a CNN}

The results of the classification of the alpha-decay events using the developed CNN model were compared to those of a former method without a CNN. In the former method, vertices and tracks are reconstructed in a three-dimensional space by combining the full information on the vertices and associated tracks from Vertex Picker \cite{YOSHIDA201786}. 
It should be noted that the method with the CNN developed in this study employs only cropped images from Vertex Picker. 
With the former method without a CNN, candidates of alpha-decay events are sorted using the information of the track multiplicity from the associated vertex and the track length.
The former method without a CNN sorted 2489 alpha-decay candidates from the total number of 46,948 events, including 201 true alpha-decay events. 
Thus, the precision and recall are 0.081 $\pm$ 0.006 (201/2489) and 0.788 $\pm$ 0.056 (201/255), respectively, where the error corresponds to $\sqrt{N}$.

The performance of the developed CNN model was evaluated by comparing it to the result of the former method without a CNN, and a comparison at the same recall value of 0.788 is summarised in Table \ref{table:comparison}. The developed method using the CNN selected 366 $\pm$ 18 alpha-decay candidates including 201 true alpha-decay events. Therefore, the precision  of the developed model with the CNN was obtained as 0.547 $\pm$ 0.025, which is larger by a factor of 6.8 $\pm$ 0.6 than that of the method without a CNN.


\begin{table}[t]
\caption{Comparison of performances between our former method without a CNN (w/o CNN) and the developed CNN method (w/ CNN) at similar recall. The CNN method improved the precision over that of the other approach by a factor of 6.8 $\pm$ 0.6.}
\label{table:comparison}
\begin{tabular}{l l l l}
\hline
Method & Precision & Recall & Number of candidates\\
\hline
w/o CNN & 0.081 $\pm$ 0.006 & 0.788 $\pm$ 0.056 & 2489 \\
w/ CNN & 0.547 $\pm$ 0.025 & 0.788   & 366 $\pm$ 18 \\
\hline
\end{tabular}
\end{table}

\section{Summary}

We developed an event classifier with a CNN for alpha-decay events recorded as a micro-graphs of vertex-like objects in nuclear emulsion. 
The developed CNN models were efficiently trained to discriminate between images of alpha-decay events and other vertex-like objects by employing random augmentation and over sampling. 
The Average Precision Score for the TEST dataset, which is a metric of the classification performance, was determined to be 0.740 $\pm$ 0.009.
The performance of the developed CNN model was compared with a former method without a CNN. The deduced precision, i.e., 0.547 $\pm$ 0.025, of the developed CNN model at a recall of 0.788 is larger than that of the former method without a CNN by a factor 6.8 $\pm$ 0.6. 
It was revealed that the load of the visual inspection will be reduced to approximately 1/7 while maintaining a similar efficiency by introducing the developed CNN method in comparison to the former method without a CNN. The developed technique described in the present paper will be a foundation for the further development to discover a number of double hypernuclei through the overall scanning of the emulsion sheets of the J-PARC E07 experiment.


\section*{Acknowledgement}

This work was supported by JSPS KAKENHI Grant Numbers 16H02180, 20H00155, and 19H05147 (Grant-in-Aid for Scientific Research on Innovative Areas 6005).
We thank the J-PARC E07 collaboration for providing the emulsion sheets.
We also thank Prof. H. Tamura at Tohoku University for the fruitful discussions.



\bibliography{mybibfile}

\end{document}